\documentstyle[preprint,aps]{revtex}
\begin{document}
\preprint{DAMTP/R-96/56}
\draft
\tighten
\title{EVOLUTION OF NEAR-EXTREMAL BLACK HOLES}

\author{S.W. Hawking\footnote{E-mail: swh1@damtp.cam.ac.uk} and
    M. M. Taylor-Robinson \footnote{E-mail:
    mmt14@damtp.cam.ac.uk}}
\address{Department of Applied Mathematics and Theoretical Physics,
\\University of Cambridge, Silver St., Cambridge. CB3 9EW}
\date{\today}
\maketitle

\begin{abstract}
Near extreme black holes can lose their charge and decay by 
the emission of massive BPS charged particles.
We calculate the greybody factors for low energy charged and neutral
scalar emission from four and five dimensional near extremal
Reissner-Nordstrom black holes. We use the corresponding 
emission rates to obtain ratios of the rates of loss of excess energy 
by charged and neutral emission, which are moduli independent,
depending only on the integral charges and the horizon potentials. 
We consider scattering experiments, finding that 
evolution towards a state in which the integral charges are equal is
favoured, but neutral emission will dominate the decay back to
extremality except when one charge is much greater than the others. 
The implications of our results for the agreement between black hole
and D-brane emission rates and for the
information loss puzzle are then discussed. 

\end{abstract}
\narrowtext

\section{Introduction}
\noindent

In the last year, there has been rapid progress in the use of D-branes
to describe and explain the properties of black holes. In a series of
papers, starting with \cite{VS},
the Bekenstein-Hawking entropies for the most general 
five dimensional BPS black holes in string theory were derived by counting
the degeneracy of BPS-saturated D-brane bound states. Later these
calculations were extended to near-extremal states \cite{SH}, 
in the particular sector of the moduli space accessible to string techniques
described by Maldacena and Strominger as the ``dilute gas'' region. 
There is some evidence, though no rigorous
derivation as yet, that the agreement can be extended throughout the moduli
space of the near-extremal black holes \cite{HMS}.

These ideas were then extended to supersymmetric four-dimensional black
holes with \linebreak
regular horizons \cite{MS2}, \cite{JKM}. In \cite{T},
\cite{KT}, \cite{GKT}, it
was argued that it is useful to view the four-dimensional black holes
as dimensionally reduced configurations of intersecting branes in 
M-theory. Such configurations again permit the derivation of the entropy of
the four-dimensional state in terms of the degeneracy of the brane
bound states.

More recently, attention has been focused on the calculation of decay
rates of five-dimensional
black holes and the corresponding D-brane configurations. It was
first pointed out in \cite{CM} that the decay rate of the D-brane 
configuration exhibits the same behaviour as that of the black hole \cite{H}, 
when we assume that the number of right moving oscillations
of the effective string is much smaller than the number of left moving
ones. In a surprising paper by Das and Mathur \cite{DM}, the numerical
coefficients were found to match and it has recently been shown \cite{MS}
that the string and semiclassical calculations also agree when we drop
the assumption on the right moving oscillations. For four dimensional 
black holes intersecting brane models of
four-dimensional black holes also give agreement between M-theory and 
semi-classical calculations of decay rates \cite{GK1}, \cite{GK}.
In the last month, a rationale for the agreement between the 
properties of near extremal D-brane and corresponding black hole states 
in the dilute gas region has been proposed \cite{Ma}. 

\bigskip

These D-brane and M-theory calculations are restricted to 
certain limited regions of the black hole parameter space. In this paper,
we calculate the semi-classical
emission rates in a sector of the moduli space which is
out of the reach of D-brane and M-theory techniques (at present). We 
then obtain moduli independent quantities describing the ratio of
charged and neutral scalar emission rates and confirm that they are in
agreement with the rates calculated in the dilute gas region of the
moduli space. Thus scattering from black holes displays
a certain universal structure for states not too far from extremality. 

One can get an idea of when charged emission will be important 
compared to neutral emission by considering the expression
for the entropy. For the five dimensional extreme black hole 
this is 
\begin{equation}
S = 2\pi \sqrt{n_{1}n_{5}n_{K}}, \label{intet}
\end{equation}
where $n_1,n_5,n_K$ are integers that give the 1 brane, 5 
brane and Kaluza-Klein charges respectively. The emission 
a massive charged BPS particle will reduce at least one of 
the integers (say $n_K$) by at least one. This will cause a 
reduction of the entropy of 
\begin{equation}
\Delta S =\sqrt{n_1n_5 \over n_K}. \label{etr}
\end{equation}
The emission of Kaluza-Klein charge will be suppressed by a factor 
of $\exp(\Delta S)$ and will be small unless 
\begin{equation}
n_K>n_1n_5.
\end{equation}
Thus it seems that charged emission will occur most readily 
for the greatest charge and will tend to equalise the charges. 
However, when the charges are nearly equal, charged emission of any 
kind will be heavily suppressed. On the other hand, neutral 
emission can take place at very low energies and so will not 
cause much reduction of entropy. One would therefore expect 
it to be limited only by phase space factors and to dominate over 
charged emission except when one charge is much greater than 
the others. The situation with four dimensional black holes
is similar except that there are four charges. Again charged 
emission will tend to equalise the charges but neutral emission will 
dominate except when one charge is much greater than the 
others. In what follows we shall consider the five dimensional 
case and treat four dimensional black holes in the appendix.

In section \ref{fds} we start by calculating the rates of emission of 
neutral and charged
scalars from near extremal five-dimensional Reissner-Nordstrom black
holes. We find that the ratio of the rates of energy loss by 
charged and neutral emission are moduli independent; they depend only
on the integral charges \footnote{We distinguish here between
 charges normalised to be integers, which we call {\it integral} 
 charges, and the {\it physical} charges, which depend also on moduli.}
 and the horizon potentials. Neutral emission
always dominates charged emission, unless one of the integral charges 
is much greater than the product of the other two. 

We then discuss the implications for scattering from the black hole; 
it was suggested in \cite{MS} that under some circumstances the
black hole will decay before we can measure its state. We point out an
error in their analysis, and show that it should be possible to obtain
entropy in the outgoing radiation equal to that of the black hole
state without the black hole decaying. 

Finally, in section \ref{con}, we discuss the implications of our results
for the information loss question. It has been explicitly shown that
the emission rates from near extremal black holes and D-branes agree
in the sectors of the moduli space accessible to string calculations.
One would expect that this agreement 
between the D-brane and black hole emission rates would continue
throughout the entire moduli space of near BPS states, although a
verification is not yet possible. Now for the
D-brane configuration we can determine the microstate when the entanglement
entropy in the radiation is equal to that of the D-brane system.
Since it is possible to obtain such an entropy in the outgoing
radiation from the black hole before it decays,
it might seem as if we can extract enough information to determine the 
black hole microstate without it decaying. That is, there would
seem to be no obstruction to
scattering radiation from the black hole and obtaining information
from the outgoing radiation.
One might then expect any further scattering to be unitary 
and predictable.

This however by no means settles the information question. 
Although scattering off a D-brane regarded as a surface in 
flat space is unitary, it is not so obvious that information 
cannot be lost if one takes account of the geometry of the 
D-brane. The causal structure may have past and future 
singular null boundaries like horizons and, as with horizons, 
there is no reason that what comes out of the past surface 
should be related to what goes into the future surface. 
In the case of a static brane of one kind, there will be no 
information loss and the scattering will be unitary because 
this corresponds under dimensional reduction to a black hole 
of zero horizon area. However, in the case of four and five 
dimensional black holes with four and three non zero charges 
respectively, the effects of the charges balance to give a
non singular horizon of finite area and one might expect non
unitary scattering with information loss. 

\section{Five dimensional scattering} \label{fds}
\noindent

In this section, following \cite{CM}, \cite{DM} and \cite{MS},
we consider scattering from a five dimensional black hole 
carrying three electric charges; such black hole states were 
first constructed in \cite{HMS} and \cite{CY}.
We will work with a near extremal solution which is a solution of the
low energy action of type IIB string theory compactified on a
torus. Then, the five-dimensional metric in the Einstein frame is:
\begin{equation}
ds^{2} = - h f^{-2/3} dt^{2} + f^{1/3} ( h^{-1} dr^{2} + r^{2} 
d\Omega_{3}^{2}),
\end{equation}
where 
\begin{equation}
h = (1 - \frac{r_{0}^{2}}{r^{2}}), \
f = (1 + \frac{r_{1}^{2}}{r^{2}})(1 + \frac{r_{5}^{2}}{r^{2}})
     (1 + \frac{r_{K}^{2}}{r^{2}}).
\end{equation}
and the parameters $r_{i}$ are related to $r_{0}$ by:
\begin{equation}
r_{1}^{2} = r_{0}^{2} \sinh^{2}\sigma_{1}, \ r_{5}^{2} = r_{0}^{2} \sinh^{2} 
\sigma_{5}, \ r_{K}^{2} = r_{0}^{2} \sinh^{2} \sigma_{K}. \label{chr}
\end{equation}
We require here only the metric in the Einstein frame; the other
fields in the solution may be found in \cite{MS}. 
The extremal limit is $r_{0} \rightarrow 0$, $\sigma_{i} \rightarrow
\infty$ with $r_{i}$ fixed; we shall be interested in the sections of
the moduli space where the
BPS state is the extreme Reissner-Nordstrom solution, where the
limiting values of $r_{i}$ are equal to $r_{e}$, the Schwarzschild radius.

We may regard the black hole as the 
compactification of a six-dimensional black string
carrying momentum about the circle direction; we will
be using this six-dimensional solution in the following sections, and
the metric (in the Einstein frame) is given by:
\begin{eqnarray}
ds^2 &=& (1 + \frac{r_1^2}{r^2})^{-1/2} ( 1 + \frac{r_5^2}{r^2})^{-1/2}
 [ - dt^2 +dx_5^2 + \frac{r_0^2}{r^2} (\cosh\sigma_{K} dt + \sinh \sigma_{K}
 dx_5)^2] \nonumber \\
 &+& ( 1 + \frac{r_1^2}{r^2})^{1/2}(1 + \frac{r_5^2}{r^2})^{1/2} 
\left[(1- \frac{r_0^2}{r^2})^{-1} dr^2 + r^2 d \Omega_3^2 \right]. \label{str}
\end{eqnarray}
We assume that we are in the very near extremal region where
$r_{0} \ll r_{e}$, and moreover will consider all three hyperbolic
angles to be finite.
It is here that our analysis differs from previous work; with this
choice of parameters, we move away from the dilute gas region and a 
straightforward D-brane analysis of emission rates is not possible. 

The entropy is:
\begin{equation}
S = \frac{A_{h}}{4 G_{5}} = \frac{2 \pi^{2} r_{0}^{3} 
\prod_{i}\cosh\sigma_{i}}{4 G_{5}} \label{ma}
\end{equation}
whilst the Hawking temperature is defined by:
\begin{equation}
T_{H} = \frac{1}{2\pi r_{0} \prod_{i} \cosh\sigma_{i}}.
\end{equation}
We may define symmetrically normalised charges by:
\begin{equation}
\frac{1}{2} r_{0}^{2} \sinh 2\sigma_{i} = Q_{i}.
\end{equation}
For simplicity of notation, we assume throughout the paper that all
charges are positive; obviously for negative charges we
simply insert appropriate moduli signs. 
Our notation for the three charges $Q_{1}$, $Q_{5}$, $Q_{K}$
indicates their origin in D-brane models, from 1D-branes, 5D-branes,
and Kaluza-Klein charges respectively. The energy in the BPS limit is:
\begin{equation}
E = \frac{\pi}{4G_{5}} [Q_{1} + Q_{5} + Q_{K}] \label{mass}
\end{equation}
where $G_{5}$ is the five dimensional Newton constant,
with the excess energy for a near extremal state being
\begin{equation}
\Delta E = \frac{\pi r_{0}^2}{4 G_{5}} \sum_{i} e^{-2\sigma_{i}}. 
\end{equation}
It was stated in \cite{HMS} that the near extremal solution is
specified by six independent parameters, which we may take to be the
mass, three charges, and two asymptotic values of scalar
fields. However, there are in fact only {\it five} independent
parameters; once we fix the three charges, as well as $r_{0}$ and one
hyperbolic angle, the other two hyperbolic angles are fixed. So we specify
the state of the black hole by its mass, three charges and only {\it
  one} extremality parameter. If the BPS state is Reissner-Nordstrom, 
then excitations away from extremality leave the geometry
Reissner-Nordstrom, since the three hyperbolic angles are the same. 
For small excitations, the relationship between the temperature and the
excess energy is 
\begin{equation}
T_{H} = \frac{2}{\pi r_{e}} \sqrt{\frac{G_{5}\Delta E}{\pi r_{e}^2}},
\label{tfe}
\end{equation}
which will be useful in the following. 
With appropriate normalisations, we can define the  
potentials associated with the charges as:
\begin{equation}
A_{i} = \frac{Q_{i} dt}{(r^{2} + r_{i}^2)},
\end{equation} 
with the potentials on the horizon $r = r_{0}$ being:
\begin{equation}
A_{i} = \frac{Q_{i} dt}{(r_{0}^{2} + r_{i}^{2})}.
\end{equation}
For perturbations which leave the
compactification geometry passive, we obtain the standard
Reissner-Nordstrom solution by the rescaling $\bar{r}^{2} = (r^{2} +
r_{i}^{2})$ which gives the solution in the familiar form:
\begin{eqnarray}
ds^{2} &=& - (1- \frac{r_{+}^{2}}{\bar{r}^2})(1- \frac{r_{-}^{2}}{\bar{r}^2})
dt^{2} + \frac{1}{(1- \frac{r_{+}^{2}}{\bar{r}^2})
(1- \frac{r_{-}^{2}}{\bar{r}^2})}d\bar{r}^{2} + \bar{r}^2
d\Omega_{3}^{2}, \nonumber \\
T_{H} &=& \frac{1}{2\pi} (\frac{r_{+}^{2} - r_{-}^{2}}{r_{+}^3}), 
\label{5rn}\\
A_{i} &=& \frac{Q dt}{\bar{r}^2}, \nonumber 
\end{eqnarray}
where in the extremal limit $r_{\pm}^{2}$ are equal to $Q$. 

\subsection{Neutral scalar emission}
\noindent

In this section we compute the absorption probability for neutral scalars
by the slightly non-extremal black hole. Our discussion parallels that
in \cite{MS}, and we hence give only a brief summary of the
calculation. We solve the Klein Gordon equation for 
a massless scalar on the fixed background; taking the field to be of
the form $\Phi = e^{-i\omega t} R(r)$, we find that:
\begin{equation}
[ \frac{h}{r^3} \frac{d}{dr} (h r^{3}  
\frac{d}{dr}) + \omega^{2} f ] R = 0. \label{kg5}
\end{equation}
where we have taken $l=0$ since we will be interested in very low
energy scalars. We assume the low energy condition:
\begin{equation}
\omega r_{e} \ll 1, \label{lec}
\end{equation}
where we treat the ratios $r_{i}/r_{e}$ as approximately one. 

Solutions to the wave equation may be approximated by matching near and 
far zone solutions. We divide the space into two regions: the far zone 
$r > r_{f}$ and the near zone $r < r_{f}$, where $r_{f}$ is the point where 
we match the solutions. $r_{f}$ is chosen so that
\begin{equation}
r_{0} \ll r_{f} \ll r_{1}, r_{5}, r_{K}, \ \omega r_{e} 
(\frac{r_{e}}{r_{f}}) \ll 1.
\end{equation}
Now in the far zone, after setting $R = r^{-3/2}\psi$ and $\rho = \omega r$, 
(\ref{kg5}) reduces to:
\begin{equation}
\frac{d^{2}\psi}{d\rho^{2}} + (1 - \frac{3}{4\rho^{2}}) \psi = 0,
\end{equation}
which has the solution for small $r$, $r \approx r_{f}$,
\begin{equation}
R = \sqrt{\frac{\pi}{2}}\omega^{3/2} [ \frac{\alpha}{2} + \frac{\beta}{\omega}
(c + \log(\omega r) - \frac{2}{\omega^{2}r^{2}})], \label{near} 
\end{equation}
where $\alpha$, $\beta$ and $c$ are integration constants, to be determined
by the matching of the solutions. The solution for large $r$ is
\begin{equation}
R = \frac{1}{r^{3/2}} [e^{i\omega r}(\frac{\alpha}{2}e^{-i3\pi/4} 
- \frac{\beta}{2}e^{-i\pi/4}) + 
e^{-i\omega r}(\frac{\alpha}{2}e^{i3\pi/4} 
- \frac{\beta}{2}e^{i\pi/4})]. \label{far}
\end{equation}
However, in the near zone, we have the equation:
\begin{equation}
\frac{h}{r^3} \frac{d}{dr} (h r^{3} \frac{dR}{dr}) 
+ [\frac{(\omega r_{1}r_{K}r_{5})^{2}}{r^{6}} + \frac{\omega^{2}( 
r_{1}^{2}r_{5}^{2} + r_{1}^{2}r_{K}^{2} + r_{5}^{2}r_{K}^{2} )}{r^4}]
R = 0.
\end{equation}
Defining the variable $v = r_{0}^{2}/r^{2}$, the equation becomes
\begin{equation}
(1 - v) \frac{d}{dv}(1-v)\frac{dR}{dv} + (D + \frac{C}{v}) R = 0, \label{de}
\end{equation} 
where
\begin{equation}
D = (\frac{\omega r_{1}r_{5}r_{K}}{2r_{0}^{2}})^{2}, 
C = ( \frac{\omega^{2}(r_{1}^{2}r_{5}^{2} + r_{1}^{2}r_{K}^{2} +
 r_{5}^{2}r_{K}^{2})}{4r_{0}^{2}}).
\end{equation}
(\ref{de}) is the same near zone equation as in \cite{MS}, but with different definitions of the quantities $C$, $D$. We can hence write down
the solution for $R$ in the near zone as:
\begin{equation}
R = A (1 - v)^{-i(a+b)/2}\frac{\Gamma(1-ia-ib)}{\Gamma(1-ia)\Gamma(1-ib)}, 
\end{equation}
with $A$ a constant to be determined and
\begin{equation}
a = \sqrt{C+D} + \sqrt{D}, \ b = \sqrt{C+D} - \sqrt{D}.
\end{equation}
By matching $R$ and $R'$ at $r = r_{f}$, we may determine the constants 
$\alpha$ and $A$, and then find the absorption probability for the
S-wave by:
\begin{equation}\sigma_{abs}^{S} = \frac{[R^{\ast}hr^{3}\frac{dR}{dr} - c.c]_{\infty}}
{[R^{\ast}hr^{3}\frac{dR}{dr} - c.c]_{r_{0}}}.
\end{equation}
That is, we take the ratio of the flux into the black hole at the horizon to 
the incoming flux from infinity. Using the values of integration constants
determined by matching, we find
\begin{equation}
\sigma_{abs}^{S} = \pi^{2} r_{0}^{2} \omega^{2} ab
\frac{(e^{2\pi(a+b)} - 1)}{(e^{2\pi a} -1)(e^{2\pi b} -1)}. \label{sabs}
\end{equation}
The values of $a$ and $b$ are:
\begin{eqnarray}
a &=& \frac{\omega}{r_{0}^{2}}(r_{1}r_{5}r_{K}), \\
b &=& \frac{\omega}{4} (\frac{r_{1}r_{5}}{r_{K}} + \frac{r_{1}r_{K}}{r_{5}}   
+\frac{r_{5}r_{K}}{r_{1}}).
\end{eqnarray}
If we now impose the conditions that the BPS state is
Reissner-Nordstrom, then for small deviations away from extremality,
\begin{equation}
a = \frac{\omega}{2\pi T_{H}}, \ b = \frac{3 \omega r_{e}}{4}.
\end{equation}
Since the Hawking temperature $T_{H}$ is much smaller than $1/r_{e}$ 
in the near extremal limit, $a \gg b$ and the low energy condition 
(\ref{lec}) implies that $b \ll 1$. From (\ref{sabs}) we find:
\begin{equation}
\sigma_{abs}^{S} =  \frac{1}{2} \pi \omega^{3} r_{1}r_{5}r_{K}  
       =  \frac{1}{4\pi} A_{h} \omega^{3}. 
\end{equation}
In fact, the low energy condition on $\omega$ implies that the
absorption cross-section exhibits the universal
behaviour discussed in \cite{DGM}; however, we will use the more 
general solutions to (\ref{kg5}) in the following sections (when we
impose different conditions on the relative sizes of the $r_{i}$). 

We can obtain the emission rate by converting the S-wave absorption
probability to the absorption cross-section using
\begin{equation}
\sigma_{abs} = \frac{4\pi}{\omega^{3}} \sigma_{abs}^{S},
\end{equation}
and then using the formula for the Hawking emission rate
\begin{equation}
\Gamma = \sigma_{abs} \frac{1}{(e^{\frac{\omega}{T_{H}}} - 1)} 
\frac{d^{4}k}{(2\pi)^{4}},
\end{equation}
to obtain
\begin{equation}
\Gamma = A_{h} \frac{1}{(e^{\frac{\omega}{T_{H}}} - 1)} 
\frac{d^{4}k}{(2\pi)^{4}}.
\end{equation}

\subsection{Charged scalar emission}
\noindent

We now turn to the problem of calculating the corresponding S-wave 
absorption cross-section for charged scalars; for simplicity, we
consider particles carrying only one type of charge.
Let us consider a scalar carrying the Kaluza-Klein charge; such a 
particle is massive in five dimensions, with its mass satisfying a BPS
bound, but in six dimensions the particle is massless, carrying
quantised momentum in the circle direction. We can hence obtain the
equation of motion by solving the massless Klein Gordon equation 
for a minimally coupled scalar in the six dimensional 
background (\ref{str}). Considering only the
S-wave component, and taking a field of the form
$\Phi = e^{-i\omega t-imx^{5}} R(r)$, we obtain the radial equation:
\begin{equation}
\frac{h}{r^3} \frac{d}{dr} (h r^{3} \frac{dR}{dr}) + 
(1 + \frac{r_{1}^{2}}{r^2})(1 + \frac{r_{5}^{2}}{r^2})
 [\omega^2 - m^{2} + (\omega \sinh \sigma_{K} - m \cosh \sigma_{K})^{2}
\frac{r_{0}^{2}}{r^2}] R = 0 \label{cr}
\end{equation}
where $m$ is the BPS mass of the particle. We obtain the same
equation, with the appropriate permutations of $r_{i}$ and
$\sigma_{i}$, for the propagation of BPS scalars carrying
charges with respect to $A_{1}$ and $A_{5}$ from the coupled
Klein-Gordon equation. 

By defining new variables, 
\begin{equation}
\omega'^{2} = \omega^{2} - m^{2}, r_{K}' = r_{0} 
\left |\sinh\sigma_{K}'\right |,
e^{\pm\sigma_{K}'} = e^{\pm\sigma_{K}}\frac{(\omega \mp m)}{\omega'},
\end{equation}
we bring the equation into the form (\ref{kg5}), and we can hence
obtain the S-wave absorption fraction from (\ref{sabs}), replacing the
variables with primed variables. 
Expressed in the primed variables, the low energy condition becomes
\begin{equation}
\omega' r_{e} \ll 1, \omega' r_{K}' \ll 1,
\end{equation}
that is, the momentum of the emitted particle
must be much smaller than the reciprocal of both the Schwarzschild radius
and the effective radius $r_{K}'$. 
In calculating the absorption probability for neutral scalars, we assumed
that in the near extremal solution the ratios $r_{e}/r_{i}$ are of order
one for each of the radii. However, 
if we rewrite $r_{K}'$ in terms of $r_{K}$, we find that:
\begin{equation}
r_{K}' = r_{K} \frac{\left |\omega - m/\phi_{K} \right|}{\omega'},
\end{equation}
where $\phi_{K} = \tanh\sigma_{K}$ is the Kaluza-Klein
electrostatic potential on the horizon. 
Let us take the low energy limit, assuming that the emitted particles
are non-relativistic, with kinetic energy $\delta$; the near extremality
condition implies that $\phi_{K} = 1 - \mu_{K}$ with $\mu_{K} \ll 1$. 
Under these conditions,
\begin{equation}
r_{K}' = r_{K} \frac{\left | \delta - m \mu_{K} \right |}{\sqrt{2m\delta}}. 
\end{equation}
There are two regions of interest. If the kinetic energy is of the
same order or greater than $m \mu_{K}^2$, then $r_{k}' \le r_{e}$, and
in solving (\ref{kg5}) we must impose this condition. As before, the
low energy condition implies that the momentum of the emitted particle
is small compared to the scale set by the Schwarzschild radius.

The other region of interest is when the kinetic energy is very small,
that is, $\delta \le m \mu_{K}^2$; we then find that $r_{K}'$ is of
the same order or greater than the Schwarzschild radius. Since the
thermal factor in the emission rate is large at small kinetic
energies, it is important to consider carefully the behaviour of the
absorption probability in this limit.
Note that in this region the enforcement of the low
energy condition requires that 
\begin{equation}
m r_{e} \ll \frac{1}{\mu_{K}}.
\end{equation}
We consider first the region where the kinetic energy is of the same
order or greater than the potential term; the solution
(\ref{sabs}) applies, using the primed variables, where 
$a$ and $b$ are determined under the condition $r_{K}' \le r_{e}$ as
\begin{eqnarray}
a &= & \frac{\omega' r_{1}r_{5}}{2 r_{0}} e^{\sigma_{K}'} 
=  \frac{(\omega - m)}{2 \pi T_{H}}, 
\\
b &=& \frac{\omega' r_{1}r_{5} }{2 r_{0}} e^{-\sigma_{K}'}
 = \frac{(\omega + m)r_{e}}{4},
\end{eqnarray}
and we assume that deviations from the extreme Reissner-Nordstrom
state are small. In addition,
\begin{eqnarray}
(a+b) &=& \frac{(\omega - m \phi_{K})}{2\pi T_{H}}, \\
 ab &=& \frac{(\omega^{2} - m^{2}) r_{e}^{4}}{4 r_{0}^{2}}, 
\end{eqnarray}
so that substituting into (\ref{sabs}) we find that
\begin{equation}
\sigma^{S}_{abs} = \frac{1}{8} A_{h} (\omega^{2} -
m^{2})^{2} r_{e} \frac{(e^{\frac{\omega - m\phi_{K}}{T_{H}}}- 1)}
{(e^{\frac{(\omega - m)}{T_{H}}} - 1)(e^{ \frac{1}{2}\pi (\omega + m) r_{e}} -
  1)}. \label{tw}
\end{equation}
This is the general expression for the absorption probability, and
applies even when the mass is of the order of $1/r_{e}$, provided that
the kinetic energy is greater than $m \mu_{K}^2$. It is interesting to
consider the limiting expression when the kinetic energy is much
smaller than the temperature. Now, the Hawking temperature is
\begin{equation}
T_{H} = \frac{\mu}{\pi r_{e}}, \label{temp}
\end{equation}
where we have used the fact that for the Reissner-Nordstrom solution 
$\mu_{i} \equiv \mu$. The condition on the kinetic
energy implies that $\delta$ is only smaller than the temperature 
when $m r_{e} \ll 1$. That
is, the mass must be small on the scale of the Schwarzschild radius. 
We can then expand out the exponentials in (\ref{tw}) to obtain  
\begin{equation}
\sigma^{S}_{abs} = \frac{1}{4\pi} A_{h} (\omega - m)(\omega + m)
(\omega - \phi_{K} m). \label{pro}
\end{equation}
The corresponding probabilities for BPS particles carrying the other
two charges are given by the same expression, with appropriate masses
and potentials. Since the horizon potentials for all three fields are the
same, under the conditions that the extreme geometry is
Reissner-Nordstrom, the probabilities for the three types of charges
differ only in the BPS masses. In the limit that the $m \mu_{K} \ll
\delta$, we find that
\begin{equation}
\sigma^{S}_{abs} = \frac{1}{4\pi} A_{h} (\omega - m)^2 (\omega + m).
\end{equation}
We now find the absorption probability in the limit that the kinetic
energy is very small, $\delta \le m \mu_{K}^2$. With these conditions,
we find that the absorption probability is given by (\ref{sabs}) with
$a$ and $b$ given by
\begin{eqnarray}
a &=& \frac{\omega' r_{e}^2 r_{K}'}{r_{0}^2}, \nonumber \\
b &=& \frac{\omega'}{4} (\frac{r_{e}^2}{r_{K}'} + 2 r_{K}').
\end{eqnarray}
Now the condition $\omega' r_{K}'$ implies that $b \ll 1$, and so we find that
\begin{equation}
\sigma^{S}_{abs} = \frac{1}{4\pi} \omega'^2 A_{h} m \mu_{K}, \label{cf}
\end{equation}
where the low energy condition implies that $m \ll 1/r_{e}\mu_{K}$.
We obtain the absorption cross-section from the S-wave absorption
probability using:
\begin{equation}
\sigma_{abs} = \frac{4\pi}{\omega'^{3}} \sigma_{abs}^{S}, 
\end{equation}
and then obtain the emission rate from the expression
\begin{equation}
\Gamma = v \sigma_{abs} \frac{1}{(e^{\frac{(\omega-m\phi_{K})}{T_{H}}} - 1)} 
\frac{d^{4}k}{(2\pi)^{4}}, \label{yr}
\end{equation}
Now from (\ref{tw}) we see that the general expression for the
emission rate (assuming that $\delta \ge m \mu_{K}^2$) is
\begin{equation}
\Gamma = \frac{\pi}{2} A_{h} (\frac{\omega^2 - m^2}{\omega}) r_{e} 
\frac{1}{(e^{\frac{(\omega - m)}{T_{H}}} - 1)}
\frac{1}{(e^{\frac{1}{2}\pi(\omega + m)r_{e}} - 1)} \frac{d^4 k}
{(2\pi)^4}. \label{gen}
\end{equation}
In the limit of small kinetic energy, we find that
\begin{equation}
\Gamma = A_{h} \mu_{K} \frac{1}{e^{\pi m r_{e}} - 1}
\frac{d^4k}{(2\pi)^4},
\end{equation}
where $m r_{e} \ll 1/\mu_{K}$. This holds not only for $\delta \ge m
\mu_{K}^2$, but also for smaller kinetic energies, since we find the
same emission rate from the absorption probability (\ref{cf}). So,
although it was important to consider carefully the behaviour of the
cross-section for very small kinetic energy, the emission rate
(\ref{gen}) in fact holds for all low energy emission.

It is interesting to look at the relative values of the neutral and
charged emission rate at very small (kinetic) energy. At small energy,
$k^3 dk = 2 m^2 \delta d\delta$, and so assuming that the mass is
small on the scale set by the Schwarzschild radius, we find 
that
\begin{equation}
\Gamma_{neut} = \frac{1}{4\pi^{2}} A_{h} T_{H} m \delta d\delta, \label{one}
\end{equation}
where we have integrated out the angular dependence. Now the emission
rate of neutral scalars at very 
low energy such that $k^{3} dk = \delta^{3} d\delta$ is
\begin{equation}
\Gamma = \frac{1}{8\pi^{2}} A_{h} T_{H} \delta^{2} d\delta,
\end{equation}
and thence the ratio of emission rates is\begin{equation}
\frac{\Gamma_{char}}{\Gamma_{neut}} = \frac{2 m}{\delta}.
\end{equation}
Since the charged particles are emitted non-relativistically, 
emission of light charged particles dominates the
emission of neutral scalars at very small energy. Since the density of
states factor in (\ref{gen}) peaks for small kinetic energy, this
indicates that the total rate of emission of light charged particles dominates
that of neutrals. When we integrate the differential emission rate for
neutrals, we find that the total rate of emission is
\begin{equation}
\Gamma^{tot}_{neut} = \frac{\pi^2}{120} A_{h} T_{H}^4.
\end{equation}
The total emission rate of light charged particles is approximated by 
\begin{equation}
carefullly\Gamma^{tot}_{char} = \frac{\zeta(3)}{2 \pi^2} A_{h} m T_{H}^3,
\end{equation}
and we find that most of the particles are emitted
with kinetic energies of the order of $m \mu_{K}$. So comparing the
total neutral and charged emission rates we find that
\begin{equation}
\frac{\Gamma^{tot}_{char}}{\Gamma^{tot}_{neut}} = \frac{60 \zeta(3)}{\pi^4}
(\frac{m}{T_{H}}).
\end{equation}
Very close to extremality, the Hawking temperature is much smaller than
the BPS masses of emitted particles, and thus emission of light
charged particles dominates. 

If we now compare the
rate of emission of higher mass particles to that of neutral scalars,
at very low kinetic energy, we find
\begin{equation}
\frac{\Gamma_{char}}{\Gamma_{neut}} = \frac{2 \pi m^2 r_{e}}{\delta}
e^{- \pi m r_{e}}.
\end{equation}
So the rate of emission of high mass particles is comparable to the 
rate of emission of neutrals only over a very small range of 
kinetic energies. The total rates of emission from the black
hole are dominated by emission of particles of higher (kinetic) energy,
and we would expect neutral emission to dominate.

This is evident from the total emission rate of higher mass particles,
which we approximate by integrating the rate (\ref{gen})
\begin{equation}
\Gamma^{tot}_{char} = \frac{\zeta(3)}{2 \pi} A_{h} r_{e} m^2 T_{H}^3 e^{-\pi m
  r_{e}}.
\end{equation}
So comparing the total neutral and charged
emission rates, for high mass particles, we find that
\begin{equation}
\frac{\Gamma^{tot}_{char}}{\Gamma^{tot}_{neut}} =
\frac{60 \zeta(3)}{\pi^3} \frac{(m r_{e})^2}{\mu_{K}} e^{-\pi m
  r_{e}}. \label{comp}
\end{equation}
Then the neutral emission rate always dominates
the charged emission rate, except at extremely low temperature. 

\bigskip

Thus, for a Reissner-Nordstrom black hole, very close to the BPS
state, we expect that the dominant decay mode is via charged emission
provided that the minimum BPS mass of the charged particles is small
on the scale of the Schwarzschild radius. Emission of charged scalars
with a mass large compared to this scale is exponentially suppressed
with respect to neutral emission. 

\bigskip

In passing we mention that 
although we have been discussing emission of particles carrying a
single type of charge the calculation applies also to BPS particles
carrying all three types of charge, such that
\begin{equation}
m = m_{1} + m_{5} + m_{K}. 
\end{equation}
The equation of the motion of the particle is the coupled Klein-Gordon
equation, where we consider coupling to all three fields. 
The emission rate is (\ref{gen}),
implying that the rate of emission of particles of the same BPS mass 
is equal, whatever the distribution of the three charges, as we would
expect for a Reissner-Nordstrom state. It might seem as though the
emission of charged particles carrying several types of charges would 
be significant in determining the total charge emission rates. However,  
as we shall see in the following section, the relationships between 
the three (quantised) BPS masses are such that at most only one type of 
charged particle can be light on the scale of the Schwarzschild
radius. 

\section{Implications for measurements} \label{iom}
\noindent

Our discussion so far has involved only the effective
five-dimensional solution, which is a solution of the low energy
action of type IIB theory compactified on a torus. 
Following the notation of \cite{MS}, we can express the energy 
of the BPS state in terms of charges normalised to be integers,
$n_{1}$, $n_{5}$ and $n_{K}$ as
\begin{equation}
E = \frac{Rn_{1}}{g} + \frac{R V n_{5}}{g} + \frac{n_{K}}{R} \label{yu}
\end{equation}
where $R$ is the circle radius, $V$ is the volume of the four torus
and $g$ is the string coupling. 
In D-brane models, the integers $n_{1}$, $n_{5}$ and $n_{K}$ are
interpreted as the number of 1D-branes wrapping the Kaluza-Klein
circle, the number of 5D-branes wrapping the five torus, and the
momentum in the circle direction respectively.
We adopt the conventions of \cite{HMS},
including $\alpha' = 1$, so that all dimensional quantities are
measured in string units and the five-dimensional Newton constant
is given in terms of the moduli by $G_{5} = \frac{\pi g^2}{4 VR}$. 
In terms of the integral charges, the entropy of the BPS state takes 
the moduli independent form (\ref{intet})
and, as we discussed in the introduction, 
this formula immediately implies that charged emission is in general
suppressed. We find precisely such suppression is implied by the rates
we have calculated.

\bigskip

We first however address an issue that we have so
far neglected. In the previous section, we have 
implicitly assumed that we can take the energy of the neutral scalar,
and the kinetic energy of the emitted scalar to be arbitrarily small
compared to all other energy scales. In \cite{MSu},
Maldacena and Susskind found that the low-lying excitations of D-brane
configuration in which the Kaluza-Klein radius is large were quantised in 
units of 
\begin{equation}
\Delta E = \frac{1}{n_{1}n_{5}R} \approx \frac{G_{5}}{r_{e}^4}.
\end{equation}
For more general conditions on the moduli, one would expect there to
be light excitations of the BPS D-brane configuration of the same scale. 
It has been suggested that the existence of such a mass gap, for which
there is no analogue for the Schwarzschild black hole, can be justified
even at the level of the classical black hole solution.

It was first pointed out in \cite{PS} that the statistical 
description of a near extremal black hole breaks down as the temperature 
approaches zero. As the heat capacity approaches one, gravitational back 
reaction must be included; the scale at which such effects become 
important is an excitation energy of $G_{5}/r_{e}^4$. This 
excitation energy is of the same order as the kinetic energy of the black hole
according to the uncertainty principle. 

In \cite{HW}, it was suggested that small perturbations about extreme 
black holes for which the entropy vanishes, but the formal temperature does
not, are protected by mass gaps which remove them from thermal contact with 
the outside world. The particular class of black holes discussed was 
electrically charged dilaton black holes in four dimensions; a parameter
$a$ describes the dilaton coupling to the gauge fields with $a=0$
describing the usual Reissner-Nordstrom solution, and $a=1$ describing
a solution of particular interest in string theory. In the
case that $a > 1$, the entropy of the extreme state vanishes, with the
formal temperature diverging; the existence of mass gaps was then suggested
to prevent radiation at the extreme. For extreme states in which the entropy 
is finite and the temperature is zero - the type of states which we are 
analysing here - there are no such objections to the black hole
absorbing or emitting arbitrarily small amounts of energy, and no such
justifications for introducing mass gaps in the classical solutions. 

In \cite{KW}, and more recently in \cite{K}, the thermal factors in 
black hole emission rates were derived taking account of self-interaction.
This approach gives the appropriate thermal factors for both the 
high energy tail of the emission spectrum of a non-extremal black hole and 
also for the emission spectrum of a very near-extremal black hole, and it
is found that they differ significantly from those in the free field limit.
There are however no physical reasons for requiring the excitation 
spectrum to be quantised in the very near extremal limit in the
semi-classical theory.

One would expect the spectrum of the classical black hole to be
continuous with arbitrarily small amounts of energy being emitted and
absorbed. In the parametrisation of the previous section, the implies
that the potentials $\mu_{i}$ are continuous and not discrete. 
Our emission rates will only be valid provided that the total
excitation energy 
above the extremal state is greater than the uncertainty in the kinetic
energy of the state according to the uncertainty principle; below this
temperature our rates should be modified in the ways suggested in
\cite{KW} and \cite{K}.

It is important to note that individual $\mu_{i}$ can correspond to 
excitation energies which are smaller than the uncertainty in the
kinetic energy provided that the total excitation energy is much
greater. This will occur if one physical charge, let us say the
Kaluza-Klein charge, is much smaller than
the other two. It may at first appear as though this implies that 
the kinetic energy of emitted scalars carrying the other two charges
must be smaller than the uncertainty in kinetic energy, and much
smaller than the temperature. However the division of the excitation
energy into three sectors is artificial in the sense that charged emission 
processes reduce the excitation energies in all three sectors. So we
should still allow for the emission of scalars with kinetic energies
up to the total excitation energy in integrating to find total
emission rates. 

\bigskip

Let us firstly assume that the BPS masses of the emitted particles are
quantised in equal units, that is, $R/g = RV/g = 1/R$; then we can 
express the mass of the extreme Reissner-Nordstrom black hole as:
\begin{equation}
E = \frac{3 n \sqrt{n}}{r_{e}}
\end{equation}
where $r_{e}$ is the Schwarzschild radius and $n \equiv n_{i}$. 
The masses of the emitted BPS charged particles are quantised as
\begin{equation}m = \frac{c \sqrt{n}}{r_{e}},
\end{equation}
with c integral. When $n$ is a large integer, the emission of all 
charged particles must be suppressed at low energy as $m r_{e} \gg 1$;
from (\ref{gen}), we find that emission of particles of minimum BPS
mass is suppressed as $e^{-\pi\sqrt{n}}$. This is precisely the 
factor we would expect; the entropy loss of the black hole when it loses a 
single particle of minimum BPS mass is $\Delta S =  \pi \sqrt{n}$
and the emission rate is suppressed as $e^{-\Delta S}$.
Under these conditions, we would expect the black hole to decay back
to extremality by emission of low energy neutral scalars, except when
the Hawking temperature is very low. 

\bigskip

There is a subtlety that we will mention briefly here and then ignore; 
if the integral charges are small, a significant fraction of the mass
of the black hole will be lost when any charged particle is emitted
and  we must be more careful about the 
thermal factor. Following \cite{K}, we find that the emission rate is
suppressed as 
\begin{equation}
\Gamma = v \sigma_{abs} e^{[S_{final} - S_{orig}]} \frac{d^4k}{(2\pi)^4}
\end{equation}
which gives an exponential factor
\begin{equation}
\Gamma \propto [e^{-2\pi n (n^{1/2} - (n-1)^{1/2})}] 
\end{equation}
where we assume that a particle of minimum BPS mass is emitted. 
So for very small $n$ it is possible that a significant fraction of
the charge of the black hole is lost as the black hole decays back
towards extremality. We shall not attempt further analysis of such
states, for which the techniques of \cite{K} would be required. 

\bigskip

If a Reissner-Nordstrom BPS state for which $n_{k} \gg n_{i}$
is slightly excited from extremality, it will decay predominantly via
emission of particles carrying the Kaluza-Klein charge, since such
particles have a mass small on the Schwarzschild radius. However, as it
decays towards a state in which $n_{k} \sim n_{i}$, the emission
starts to be suppressed by the unfavourable entropy loss from the
black hole when each unit of charge is lost (\ref{etr}). This
behaviour depends only on the integral charges. For the analysis
in the dilute gas region of \cite{MS} factors of 
$e^{-R T_{L}}$ appear in the rates, where $T_{L}$ is the temperature
of the left-moving excitations of the effective string.
Since $R T_{L} \sim \Delta S$, this is precisely the
behaviour we would expect.

The authors of \cite{MS} suggested that the Kaluza-Klein charge of the
hole could be lost before the entropy in the emitted radiation was
sufficient to determine the state of the hole, but their analysis 
failed to take note of the fact that as the Kaluza-Klein charge is
reduced, further emission is suppressed. Expressed in terms of the
temperature of the left moving excitations, even if this
temperature is initially large compared to the scale set by the
Kaluza-Klein radius, it is reduced by the emission. As the temperature
approaches $1/R$, further charged emission is suppressed.
More generally, what we would
expect to happen is that the black hole evolves towards a state in
which all three charges are comparable. Thereafter, emission of even the
lightest charged state will be exponentially suppressed relative to
neutral emission. 

\bigskip

For the Reissner-Nordstrom state in which all the integral charges 
are equal, as the black hole decays back towards extremality by neutral
emission, the Hawking temperature decreases and the rate at which
the excess energy is lost by the hole becomes very small. So, very close to
extremality, the rates of loss by neutral and charged emission 
may become comparable despite the entropy loss involved in charged
emission. This is apparent from looking at ratio of the emission rates 
in (\ref{comp}) but for later convenience we compare
instead the approximate rates of energy loss for neutrals
\begin{equation}
\frac{d\Delta E}{dt}_{neut} \approx \int \Gamma \omega,
\end{equation}
with the corresponding rate for charged particles
\begin{equation}
\frac{d\Delta E}{dt}_{char} \approx \int \Gamma (\omega - m), \label{xx}
\end{equation}
where we will assume only particles of minimum BPS mass are emitted. 
Now the energy loss rate by neutral emission is
\begin{equation}
\frac{d\Delta E}{dt}_{neut} \approx \frac{3 \zeta(5)}{\pi^2} A_{h}
T_{H}^5.
\end{equation}
For a Reissner-Nordstrom state with the integral charges equal we find that:
\begin{equation}
\frac{d\Delta E}{dt}_{char} \approx \frac{\pi^3}{60} A_{h} T_{H}^4
  (\frac{n}{r_{e}}) e^{-\pi \sqrt{n}}, 
\end{equation}
and so we find the ratio of energy loss rates to be using (\ref{temp})
\begin{equation}
\frac{\frac{d\Delta E}{dt}_{char}}{ \frac{d\Delta
    E}{dt}_{neut}} \approx \frac{\pi^6}{180 \zeta(5)}
\frac{n}{\mu} e^{-\pi \sqrt{n}}. \label{rat}
\end{equation}
where $\mu$ is the deviation of the potential from one on
the horizon, and the rate of loss of all three charges is the same. 

\bigskip

The relative rates of energy loss are independent of the values
 of the moduli. Using the results of
\cite{MS}, obtained under the condition that the momentum modes are light, 
setting the charges to be equal, we find that the rates
of loss of energy by emission of KK charged particles and neutrals
compare as
\begin{equation}
\frac{ \frac{d\Delta E}{dt}_{KK}}{ \frac{d\Delta
    E}{dt}_{neut}} \approx \frac{\pi^6}{180 \zeta(5)}
 \frac{n}{\mu_{K}} e^{- \pi \sqrt{n}}.
\end{equation}
where $\mu_{K}$ is the Kaluza-Klein potential on the horizon. That is,
we find the same ratio for Kaluza-Klein charged and neutral emission
in this sector of the moduli space, confirming the modular independence
of the result. 

However, in \cite{MS}, it was assumed that only $\mu_{K}$ was non-zero,
which would imply that only Kaluza-Klein charge is lost. Under the
condition that $Q_{K}$ is much smaller than the other two charges,
this is a reasonable approximation, since (\ref{chr}) implies that 
$\mu_{K}$ is much larger than the other $\mu_{i}$ for any given 
$r_{0}$. When the integral charges are equal, then from (\ref{chr})
and (\ref{ma}), assuming that $V = 1$, we find that
\begin{equation}
\frac{\mu_{K}}{\mu_{1}} = \frac{R^2}{g},
\end{equation} 
and so $\mu_{1} (= \mu_{5})$ is much smaller than $\mu_{K}$ under these
conditions on the moduli. 

From the point of view of the five parameter classical black hole 
solution, it is not consistent to set $\mu_{i} \equiv 0$ when $r_{0}
\neq 0$, as we pointed out above. In fact, for $n_{K} \gg n_i$, the
physical charges can be comparable, and we would expect the
non-extremality parameters in each sector to be comparable also. 
Since the authors of \cite{MS} imposed the condition 
that $Q_{K} \ll Q_{i}$, even for $n_{K} \gg
n_{1}n_{5}$ (corresponding to their condition $RT_{L} \gg 1$), their
calculations are unaffected by taking $\mu_{i}$ to be finite. That is,  
$\mu_{K}$ will always be much greater than $\mu_{i}$ and for most 
purposes we can set $\mu_{i}$ to zero, although the deviation of
$\mu_{i}$ from zero in the black hole solution 
is significant, as we shall see below. 

\bigskip

It is not difficult to extend the analysis of the section above to show 
that, under the conditions $R^2 \gg g$ and $n_{i} \equiv n$, for emission 
of the other two types of charges, the energy loss rates compare as
\begin{equation}
\frac{ \frac{d\Delta E}{dt}_{i}}{ \frac{d\Delta
    E}{dt}_{neut}} \approx \frac{\pi^6}{180 \zeta(5)} 
\frac{n}{\mu_{i}} e^{- \pi \sqrt{n}},
\end{equation}
where we assume that the $\mu_{i}$ are very small, but non-zero. 
There are two important points to notice. Firstly, this is the same
ratio as we get in the Reissner-Nordstrom sector of the moduli space
above. If we assume that $\mu_{i} \equiv 0$ in this sector of the moduli
space, the ratios are not moduli independent. 
Secondly, with these conditions on the moduli, we expect that 
$\mu_{i}$ is smaller for the heavier modes. So the rate of loss of
energy by the heavier particles is actually greater, and will dominate
emission by Kaluza-Klein charged particles
\begin{equation}
\frac{ \frac{d\Delta E}{dt}_{KK}}{ \frac{d\Delta
    E}{dt}_{1}} \approx \frac{\mu_{1}}{\mu_{K}} = \frac{g}{R^2} \ll 1.
\end{equation}
The black hole loses the same
amount of entropy in emitting a unit of each charge, so the
exponential suppression factor is the same, but the physical
Kaluza-Klein charge is smaller, and is less likely to be reduced. 
 
\bigskip

For general integral charges the relative rates of loss of energy are
\begin{equation}
\frac{ \frac{d\Delta E}{dt}_{KK}}{ \frac{d\Delta
    E}{dt}_{neut}} \approx \frac{\pi^6}{180 \zeta(5)} 
\frac{n_{1}n_{5}}{n_{K} \mu_{K}} e^{- \pi\sqrt{\frac{n_{1}n_{5}}{n_{K}}}}, 
\end{equation}
with the ratio for emission of the other two particles being 
given by the same expression with appropriate permutations of indices.
If $n_{K} \ll n_{i}$, then we see that loss of 
the Kaluza-Klein charge is suppressed, and that the rate of 
loss of the other two charges dominates the neutral emission
rate at higher temperature. So decay towards a state in which
the $n_{i}$ are equal is indeed favoured although the rate of loss of charge
will be slow compared to the loss of neutrals except at low temperature.

If $n_{K} \gg n_{1}n_{5}$ we must allow for emission of 
particles of greater than the minimum BPS mass.
For a Reissner-Nordstrom solution, the mass of
Kaluza-Klein charged particles is quantised as
\begin{equation}
m = \frac{c}{r_{e}} \sqrt{\frac{n_{1}n_{5}}{n_{K}}},
\end{equation}
where $c$ is an integer; the mass is small on the scale of the
Schwarzschild radius, and charged emission will dominate neutral
emission. We calculate the rate of energy emission for a particle of
general mass $m$, using (\ref{yr}) and (\ref{xx}) as,
\begin{equation}
\frac{d\Delta E}{dt}_{char} \approx \frac{\pi^3}{60} A_{h} T_{H}^4 
\frac{m^2 r_{e}}{(e^{\pi m r_{e}} - 1 )},
\end{equation}
and integrate over all masses to find that
\begin{equation}
\frac{d\Delta E}{dt}_{KK} \approx \frac{\zeta(3)}{30} A_{h} T_{H}^4 
\frac{1}{r_{e}} \sqrt{\frac{n_{K}}{n_{1}n_{5}}}.
\end{equation}
Comparing this to the energy loss by neutral emission we find that
\begin{equation}
\frac{ \frac{d\Delta E}{dt}_{KK}}{\frac{d\Delta
    E}{dt}_{neut}} \approx \frac{\pi^3\zeta(3)}{90 \zeta(5)}
 \sqrt{\frac{n_{K}}{n_{1}n_{5}}} \frac{1}{\mu_{K}},
\end{equation}
which is the same ratio as we obtain from \cite{MS}. Thus we find
that emission of KK charged scalars dominates neutral emission,
independently of the moduli, for any near extremal state with $n_{K}$
very large.  

\bigskip

Thus we find that, 
although the absolute rates of energy emission by the black hole
are moduli dependent, the relative rates of neutral and charged
emission depend only on the integral charges and horizon
potentials. It is straightforward to demonstrate this explicitly by
extending the scattering calculations to the most general near
extremal black holes. So the scattering rates from black holes
exhibit a certain universality which follows from the modular
independence of the BPS entropy. Let us assume that the agreement
between D-brane and black hole emission rates extends throughout the
moduli space of near extremal states and
then consider the implications of our results for  
scattering experiments. 

Suppose that we excite a BPS state slightly above extremality
with low energy radiation and measure the outgoing radiation resulting
from the decay. Whatever the value of the moduli, the black hole will
decay towards a state in which the integral charges are equal,
but such a decay will only proceed rapidly if one charge is much greater than
the other two. Under the latter conditions, the 
black hole will lose a significant fraction of its charge before there
is enough information in the outgoing radiation to measure its state.
It is simple to show that, as the black hole decays from a state
with $n_{K} \gg n_{1}n_{5}$ towards a state in which the charges are
comparable, the entropy in the outgoing charged radiation is given by
\begin{equation}
\frac{\delta S_{out}}{S_{BH}} \sim \frac{1}{n_{1}n_{5}}.
\end{equation}
Now in the string picture one can measure the state of the black hole
once the entanglement entropy in the outgoing radiation is equal to
that of the black hole. 
So here the entropy in the outgoing radiation is certainly insufficient to
determine the initial state of the black hole, and the state changes
before we can measure it. However, as the black hole decays towards a 
more stable charge configuration, we might hope to be able to measure
the state after neutral emission starts to dominate.

Suppose that we start with an extreme Reissner-Nordstrom state in which
all the integral charges are equal; the maximum excitation energy we can add
and still leave the black hole in a near extremal state is 
$\Delta E \sim \sqrt{n}/r_{e}$, i.e. an energy equal to that of
the minimally charged BPS particle. We can then estimate the total 
amount of entropy in the outgoing neutral radiation as
\begin{equation}
\delta S_{out} \sim \int^{\sqrt{n}/r_{e}}_{0} \frac{d(\Delta E)}{T_{H}},
\end{equation}
and, from (\ref{tfe}), expressing the temperature as a function of 
the excess energy, we find that $\delta S_{out} \sim {n}$. 
So in order to obtain an entropy in the outgoing radiation equal to
that of
the black hole we will need of the order of $\sqrt{n}$ experiments. In
fact, for general charges and moduli, we can show that the maximum amount of 
entropy in the outgoing radiation is
\begin{equation}
\frac{\delta S_{out}}{S_{BH}} = [\frac{1}{E_{1}} + \frac{1}{E_{5}} +
\frac{1}{E_{K}}]^{1/2} \Delta E^{1/2}_{max},
\end{equation}
where the energy of the BPS state is $E = \sum_{i} E_{i}$. Since by 
definition very close to extremality the excitation energy is much
less than the smallest of the $E_{i}$, a large number of experiments 
will be required.
After these experiments we let the black hole decay right back to
the BPS state, which takes an infinitely long time. As the temperature
becomes very small, charged emission dominates neutral emission, and
we might expect the final excess energy of the black hole to be
emitted in the form of charged radiation. 

Working in the Reissner-Nordstrom sector, neutral emission
will dominate until the ratio of rates in (\ref{rat}) is approximately
one; but when this happens, the remaining excess energy is
\begin{equation}
\Delta E \sim \frac{n^{5/2} e^{- 2 \pi \sqrt{n}}}{r_{e}},
\end{equation} 
which compares to an energy scale set by the uncertainty principle of
\begin{equation}
E_{uncert} \sim \frac{1}{n^{3/2} r_{e}},
\end{equation}
which is much larger. That is, before charged emission can become significant,
the excess energy falls below the uncertainty in energy of the BPS
state (and the statistical approximations implicit in our rates break
down).  

\bigskip

We now suggest a resolution to a paradox discussed in \cite{MS}. 
If we have a
state for which the momentum modes are light, and all three integral charges
charges are comparable, then emission of any charge is suppressed. So
we might expect that we could excite the black hole by an energy
$\Delta E \gg n/R$, still remaining in the near extremal state since
the Kaluza-Klein radius is taken to be large. 
Neutral emission will dominate the decay, and
the entropy in the outgoing radiation is  
\begin{equation}
\delta S_{out} \sim n \sqrt{R \Delta E} \gg n^{3/2}.
\end{equation}
That is, the entropy in the outgoing radiation is greater than that of
the black hole, which presents a contradiction in the string picture.

However, if we attempt to 
excite the black hole with such a large excitation energy, 
we find that $r_{K} \ll r_{0}$, and $r_{1} \sim r_{0}$, where we use 
the relationship between the extremality parameters. This
implies that the black hole is very non-extremal, and its decay lies
outside the range of the near-extremal calculations. 
For a near extremal solution, we require $r_{1}, r_{5} \gg r_{0}$, and hence
we must restrict our excitation energies to
$\Delta E \le 1/R$. Under this condition, $\delta S_{out} \sim n$ 
is the maximum amount of entropy in the outgoing radiation, much smaller
than the entropy of the black hole, as required. So it is important to
take account of all three potentials; it is straightforward
to show that the near extremal calculations are valid only when the
largest of the $\mu_{i}$ is less than or of the order of $1/n_{i}$.

Scattering from analogous four dimensional
black holes carrying four $U(1)$ charges is also found to exhibit 
a universal structure which is implied by the modular independence of the
BPS entropy. The analysis differs little from that in the five
dimensional system and we include a brief summary in the appendix. 

\section{Conclusions} \label{con}
\noindent

We have shown that by repeated scattering from the black hole we can
obtain an entropy in the outgoing radiation equal to that of the black
hole before the BPS state changes. By careful 
experimentation we might then think that information about the actual
microstate could be deduced from the absorption/scattering process.
However we must be more careful about extrapolating from the D-brane
limit of the moduli space in which 
\begin{equation}
g n_{1} < 1; \ g n_{5} < 1; \ g^2 n_{K} < 1,
\end{equation}
to the black hole limit in which 
\begin{equation}
g n_{1} > 1; \ g n_{5} > 1; \ g^2 n_{K} > 1.
\end{equation}
In the former case, we have a discrete excitation spectrum.
We can use D-brane models of near extremal black holes to
describe excitations in the ``dilute gas'' region of the moduli space;
that is, we consider states in which the 
Kaluza-Klein radius is very large and the physical Kaluza-Klein charge is 
small. In this region we can describe the near extremal state in terms of 
excitation modes of an effective string of length $Rn_{1}n_{5}$ with the
excitation energy being quantised in units of the reciprocal of the
length. For a large black hole
solution in which the Kaluza-Klein radius is large then 
in terms of the non-extremality parameters
of the black hole solution, the excitation energy in the Kaluza-Klein
sector is 
\begin{equation}
\Delta E_{KK} = \frac{\pi r_{0}^2}{4 G_{5}} e^{-2\sigma_{K}} = \frac{n_{K}
\mu_{K}^2}{R},
\end{equation}
where $\mu_{K}$ is a continuous parameter. However, we will also have
non-zero excitation energies in the other two sectors, which,
assuming for simplicity that $V = 1$, are given by
\begin{equation}
\Delta E_{1} = \Delta E_{5} = \frac{gn_{K}}{R^2n_{1}} \Delta E_{KK}.
\end{equation}
In the limit that $r_{K} \ll r_{1}$, then $\Delta E_{1} \ll \Delta E_{KK}$
and by taking the radius to be sufficiently large we can choose $\Delta E_{1}
< 1/{n_{1}n_{5}R}$. For the classical solution, this means that the 
excitation energy in this sector
is smaller than the scale set by the uncertainty principle, but is
still finite because we have taken the non-extremality parameter to be
continuous. 

What this implies physically is that the large black hole can emit BPS
charged particles with all three types of charges provided that the 
temperature is finite. According to the D-brane calculations, 
only BPS particles carrying the Kaluza-Klein charge can be
emitted. That is, the
agreement between the D-brane and black hole emission rates breaks
down when the excitation energy of the near extremal black hole is
very small in one of the sectors. This will be particularly
significant if, for example, $n_{1} \gg n_{5} n_{K}$ and $\Delta E_{1}$
in the black hole solution is smaller than the uncertainty energy. 
Then according to the black
hole calculations, the dominant decay mode should be via emission of
particles carrying this charge whereas according to the D-brane model
no such emission is possible. 

Since from general duality arguments we would expect the agreement
between black hole and D-brane emission rates to hold throughout the
moduli space, the interpretation we give to this disagreement is that
in this limit the effective string model breaks down. In terms of the
moduli space analysis proposed recently in \cite{Ma}, the probability
for the system to wander into the vector moduli space, corresponding
to D-brane emission, becomes significant in this limit. Of course
as the physical charges in the BPS state become comparable, the effective
string approximation certainly breaks down; we require the D-brane model 
to describe emission of all three charges.

\bigskip

Now in the D-brane limit, if we do scattering experiments we will
indeed know in which microstate the branes are. Suppose we examine
the absorption of a (neutral) scalar of energy $2c/n_{1}n_{5}R$ by a D-brane
configuration whose excitations are described by those of an effective
string of length $n_{1}n_{5}R$. The absorption creates a pair of open
strings moving on the string and the absorption probability depends 
on the quantum microstate of the configuration. More generally, there
will be many distinct types of
excitations of the BPS state which can be interpreted in terms of, for
example, brane/anti-brane pairs, and the absorption spectrum 
will depend on the moduli and charges of the BPS state. 

Then we can see that
repeated absorption/emission processes will give us information
about the microstate. In the black hole limit, if the spectrum is continuous,
repeated scattering
processes will simply produce an ever-increasing amount of entropy in
the outgoing radiation which does not encode the state of the black
hole. Another way of describing this would be to say that classical large black
holes behave as complex extended objects with a continuous spectrum of
low-lying excitations whereas in the D-brane limit the system behaves
as an elementary particle with discrete excitation levels. 

This picture was suggested in \cite{M} where the excitation
spectrum of an isolated D-string was shown to change from one with
discrete levels to one that has no sharp levels as we go towards the
an appropriate black hole type limit. This is what we
have assumed in taking the $\mu_{i}$ to be continuous parameters, and 
such a spectrum change is of course implicit in the picture of black hole
to D-brane transition discussed in \cite{Hor}.
It would be interesting if this change of spectrum could be
demonstrated explicitly for bound states of D-branes. 

\bigskip

The aim of this paper was to attempt to reconcile the 
non unitary behaviour of black holes with the unitary behaviour of 
the corresponding D-brane configuration by demonstrating that systems
decay before one can measure their states. We have however found that this is 
not the case. Since this work was completed, a 
correspondence principle between black holes and strings has been proposed
\cite{HP} which highlights the apparent contradictions between the pictures
still further. 
There have been suggestions that information may be lost from the D-brane
configuration in subtle ways, such as 
by recoil effects involved in scattering \cite{EMN}. However, 
as we discussed in the 
introduction, we believe that if information is lost it is because
one cannot neglect the causal structure and treat the system as though it
is in flat space. This is a subject to which we hope to return in the near
future. 

\appendix
\section*{Scattering from four dimensional black holes} \label{app}
\noindent

In this appendix we show that the same modular independence of ratios of
scattering rates is found in analogous four dimensional black hole
systems. Following \cite{GK1} and \cite{GK},
we consider a four dimensional black hole with four $U(1)$ charges 
described by the metric:
\begin{equation}
ds_{4}^{2} = - F^{-1/2}H dt^{2} + F^{1/2} (H^{-1} dr^{2} + r^{2} d\Omega^{2})
\end{equation}
with
\begin{equation}
H = (1 - \frac{r_{0}}{r}), \ F = (1 + \frac{r_{1}}{r})(1 + \frac{r_{2}}{r})(1 +
\frac{r_{3}}{r})(1 + \frac{r_{4}}{r}),
\end{equation}
where for each $r_{i}$
\begin{equation}
r_{i} = r_{0} \sinh^{2}\sigma_{i}
\end{equation} 
with the $r_{0}$ and $\sigma_{i}$ being extremality parameters, such that
in the BPS limit $r_{0} \rightarrow 0$ and $\sigma_{i} \rightarrow
\infty$ with $r_{i}$ fixed. The physical charges are given by 
$Q_{i} = r_{0} \sinh\sigma_{i} \cosh\sigma_{i}$ and the energy is
\begin{equation}
E = \frac{1}{4 G_{4}} \sum_{i} Q_{i} + \frac{r_{0}}{4 G_{4}}
\sum_{i} e^{-2\sigma_{i}},
\end{equation} 
where $G_{4}$ is the four-dimensional Newton constant. The solution
may be described by six independent parameters - the mass, four
charges and one non-extremality parameter. 

In \cite{GK1} and \cite{GK}, scalar emission rates were calculated for
this metric using semi-classical and effective string model approaches
in the limit that the Kaluza-Klein parameter $r_{4}$ was much smaller
than the other $r_{i}$. It is straightforward to extend their
semiclassical calculations to general near extremal black hole
solutions for which $r_{0} \ll r_{i}$
and we do not repeat the details of the scattering
calculation here. We find that the neutral scalar emission rate 
at low energies is 
\begin{equation}
\Gamma = A_{h} \frac{1}{(e^{\frac{\omega}{T_{H}}} - 1)} 
\frac{d^3 k}{(2\pi)^3},
\end{equation}
with the emission rate of Kaluza-Klein charged scalars of mass $m$ being
\begin{equation}
\Gamma = \pi A_{h} \frac{(\omega^2 - m^2)^{3/2}}{m (\omega - m
\phi_{4})} \sqrt{\frac{r_{1}r_{2}r_{3}}{r_{4}}}
\frac{1}{(e^{2\pi m  \sqrt{\frac{r_{1}r_{2}r_{3}}{r_{4}}}} - 1)} 
\frac{1}{(e^{\frac{\omega - m}{T_{H}}} - 1)} \frac{d^3k}{(2\pi)^3},
\label{crh}
\end{equation}
where $\phi_{4}$ is the Kaluza-Klein potential on the horizon
and corresponding expressions hold for the emission of the other
charges. These rates are valid provided that the total excitation
energy is greater than the uncertainty energy, which in four
dimensions is $G_{4}/r_{e}^3$, below which scale the statistical
assumptions break down. We can express the physical charges $Q_{i}$ in terms
of moduli and integral charges $n_{i}$ as
\begin{eqnarray}
Q_{1} = \frac{n_{1}}{L_{6}L_{7}} (\frac{\kappa_{11}}{4\pi})^{2/3},
 & \ & Q_{2} = \frac{n_{2}}{L_{4}L_{5}}
 (\frac{\kappa_{11}}{4\pi})^{2/3}, \nonumber \\
Q_{3} = \frac{n_{3}}{L_{2}L_{3}} (\frac{\kappa_{11}}{4\pi})^{2/3},
& \ & Q_{4} = 8 \pi G_{4} (\frac{n_{4}}{L_1}),
\end{eqnarray}
where $L_{i}$ is the length of the $i$th internal circle, and 
$\kappa_{11}^{2} = 8 \pi G_{4} \prod_{i}L_{i}$. Such integral charges 
arise from the toroidal compactification of an eleven-dimensional 
solution, and can be interpreted in terms of intersecting brane 
representations in M-theory \cite{KT}, \cite{GKT}. $Q_{4}$ is the
Kaluza-Klein charge, deriving from the quantised momentum in a
circle direction.

The entropy of the BPS state takes the modular independent form
\begin{equation}
S = 2 \pi \prod_{i} \sqrt{n_{i}},
\end{equation}
and so we would expect charged emission to be suppressed as the 
entropy loss in emitting one unit of Kaluza-Klein charge is
\begin{equation}
\Delta S = \pi \sqrt{\frac{n_{1}n_{2}n_{3}}{n_{4}}}, \label{set}
\end{equation}
which is generally large. Exponential suppression by
precisely this factor is implied in (\ref{crh}), since
the BPS masses of particles carrying the Kaluza-Klein charge are
quantised in units of $2\pi/L_{1}$ and
\begin{equation}
\sqrt{\frac{r_{1}r_{2}r_{3}}{r_{4}}} =
\sqrt{\frac{n_{1}n_{2}n_{3}}{n_{4}}} (\frac{L_{1}}{4 \pi}).
\end{equation}
We find that the rate of energy loss by neutral emission is
\begin{equation}
\frac{d\Delta E}{dt}_{neut} \approx \frac{\pi^2}{30} A_{h} T_{H}^4,
\end{equation}
which has a different temperature dependence to the five dimensional 
expression. The rate of energy loss by Kaluza-Klein charged emission is
\begin{equation}
\frac{d\Delta E}{dt}_{4} \approx \frac{\pi^3}{15} A_{h} T_{H}^4
\frac{1}{\mu_{4}} \sqrt{\frac{n_{1}n_{2}n_{3}}{n_{4}}} 
e^{-\pi\sqrt{\frac{n_{1}n_{2}n_{3}}{n_{4}}}},
\end{equation}
where $\mu_{4}$ is the deviation from one of the Kaluza-Klein
potential on the horizon. Note the higher exponential suppression than
in five dimensions, deriving from the expression for the entropy. 
If $n_{4}$ is much greater than the product of the other three
charges, we must allow for emission of not only minimally charged
particles and 
\begin{equation}
\frac{d\Delta E}{dt}_{4} \approx \frac{\pi^3}{90} A_{h} T_{H}^4
\frac{1}{\mu_{4}} \sqrt{\frac{n_{1}n_{2}n_{3}}{n_{4}}}.
\end{equation}
Then the modular independent ratios of energy loss rates are
\begin{equation}
\frac{\frac{d\Delta E}{dt}_{4}}{\frac{d\Delta E}{dt}_{neut}}
\approx \frac{2\pi}{\mu_{4}} \sqrt{\frac{n_{1}n_{2}n_{3}}{n_{4}}} e^{- \pi
 \sqrt{\frac{n_{1}n_{2}n_{3}}{n_{4}}}}, 
\end{equation}
except for $n_{4} \gg n_{1}n_{2}n_{3}$ when
\begin{equation}
\frac{\frac{d\Delta E}{dt}_{4}}{\frac{d\Delta E}{dt}_{neut}}
\approx \frac{\pi}{3 \mu_{4}} \sqrt{\frac{n_{4}}{n_1 n_2 n_3}},
\end{equation}
which is very large close to extremality.
Our modular independent ratios are in
agreement with those derived from the emission rates in
\cite{GK1} and \cite{GK}. 
So, unless one integral charge is much greater than
the product of the other three, charged emission does not play a role
in the decay and measurements of the microstate by repeated scattering
processes appear to be feasible.

\end{document}